\documentclass[epj,twocolumn]{webofc}
\usepackage[varg]{txfonts}   
\newcommand{\ahvp}{a_\mu^{\rm hvp}}
\newcommand{\ahlbl}{a_\mu^{\rm hlbl}}
\newcommand{\be}{\begin{equation}}
\newcommand{\ee}{\end{equation}}
\newcommand{\bea}{\begin{eqnarray}}
\newcommand{\eea}{\end{eqnarray}}
\newcommand{\bd}{\begin{displaymath}}
\newcommand{\ed}{\end{displaymath}}
\newcommand{\eq}[1]{Eq.\,(\ref{#1})}
\newcommand{\Nf}{N_{\rm f}}
\newcommand{\rme}{{\rm e}}

\woctitle{Flavour changing and conserving processes}
\begin{document}
\title{Lattice calculation of the hadronic leading order contribution
  to the muon $g-2$}
%
%

\author{Hartmut Wittig\inst{1,2}\fnsep\thanks{\email{hartmut.wittig@uni-mainz.de}}
\and Antoine G\'erardin\inst{3}
\and Marco C\`e\inst{2}
\and Georg von Hippel\inst{1}
\and Ben H\"orz\inst{4}
\and Harvey B. Meyer\inst{1,2}
\and Kohtaroh Miura\inst{2}
\and Daniel Mohler\inst{2}
\and Konstantin Ottnad\inst{1}
\and Andreas Risch\inst{1}
\and Teseo San Jos\'e\inst{1,2}
\and Jonas Wilhelm\inst{1}
}

\institute{PRISMA$^+$ Cluster of Excellence and Institute for Nuclear
  Physics, University of Mainz, D-55099 Mainz, Germany
\and
  Helmholtz Institute Mainz, D-55099 Mainz, Germany
\and
  John von Neumann Institute for Computing, DESY, Platanenallee 6,
  D-15738 Zeuthen, Germany
\and
  Nuclear Science Division, Lawrence Berkeley National Laboratory,
  Berkeley, CA\,94720, USA
}

\abstract{The persistent discrepancy of about 3.5 standard deviations
  between the experimental measurement and the Standard Model
  prediction for the muon anomalous magnetic moment, $a_\mu$, is one
  of the most promising hints for the possible existence of new
  physics. Here we report on our lattice QCD calculation of the
  hadronic vacuum polarisation contribution $\ahvp$, based on gauge
  ensembles with $\Nf=2+1$ flavours of O($a$) improved Wilson
  quarks. We address the conceptual and numerical challenges that one
  encounters along the way to a sub-percent determination of the
  hadronic vacuum polarisation contribution. The current status of
  lattice calculations of $\ahvp$ is presented by performing a
  detailed comparison with the results from other groups.}
\maketitle
\section{Introduction} \label{sec:s1intro}

The anomalous magnetic moment of the muon,
$a_\mu$, has been measured by the E821 experiment
at BNL \cite{Bennett:2006fi} with a total precision of 0.54 ppm, i.e.
\be
   a_\mu\equiv{\textstyle\frac{1}{2}}(g-2)_\mu = 
   116\,592\,080(54)(33)\cdot 10^{-11}.
\ee
While the theoretical prediction based on the Standard Model (SM) is
known at the same level of precision, it differs by 3.5 standard
deviations \cite{KnechtFCCP19}. This discrepancy is currently one of
the most intriguing hints for the possible existence of physics beyond
the SM. While the absolute value of the theoretical estimate of
$a_\mu$ is dominated by the contributions from QED, with only a small
fraction coming from strong and weak interaction effects, the
situation is completely reversed regarding the assigned uncertainty:
Indeed, the dominant errors arise from the strong interaction, in
particular from the hadronic vacuum polarisation (HVP) and hadronic
light-by-light scattering (HLbL) contributions, $\ahvp$ and $\ahlbl$,
respectively. New experiments, which are designed to reduce the error
even further, are either being prepared (E34 at J-PARC
\cite{Abe:2019thb}) or have started data taking already (E989 at
Fermilab \cite{Grange:2015fou}). Since the design sensitivity of both
experiments exceeds the theoretical uncertainty by far, it is of
utmost importance to achieve a significant reduction of the latter in
the coming years.

The conventional method to determine the HVP contribution, which
enters the current SM estimate, proceeds by expressing $\ahvp$ in
terms of a dispersion integral over the hadronic cross section ratio
$R^{\rm had}(s)=\sigma(e^+ e^-\to\rm hadrons)/\sigma(e^+
e^-\to\mu^+\mu^-)$, multiplied by a kernel function
\cite{Hagiwara:2011af, Davier:2017zfy, Keshavarzi:2018mgv,
  Davier:2019can}. In the low-energy regime one relies on experimental
data for the $R$-ratio $R^{\rm had}(s)$, which implies that the SM
prediction is subject to experimental uncertainties. While a
dispersive framework to address $\ahlbl$ is being developed
\cite{Colangelo:2014dfa, Colangelo:2014pva, Colangelo:2015ama,
  Colangelo:2017qdm, Colangelo:2017fiz, Pauk:2014rfa,
  Pascalutsa:2012pr, Green:2015sra, Gerardin:2017ryf,
  Danilkin:2016hnh, Hagelstein:2017obr}, the current SM prediction is
still largely based on hadronic models \cite{Nyffeler:2016gnb}.

Simulations of QCD on a space-time lattice have emerged as a versatile
tool for the treatment of the strong interaction in the
non-perturbative regime from first principles. Lattice QCD provides
accurate estimates for SM parameters (strong coupling $\alpha_s$,
quark masses), decay constants and form factors relevant for flavour
physics, quantities describing structural properties of the nucleon
and many other observables of phenomenological relevance
\cite{Aoki:2019cca}. Recent years have also seen enormous progress
regarding calculations of the hadronic contributions to the muon $g-2$
(see Ref. \cite{Meyer:2018til} for a review).

A first-principles approach to the muon $g-2$ based on lattice QCD
avoids both the reliance on experimental data, as well as model
estimates. However, in order to have an impact on tests of the SM,
lattice QCD must deliver results for $\ahvp$ with a total error well
below 1\% and a determination of $\ahlbl$ at the $10-15$\%
level. Whether or not this can be achieved is also the subject of a
planned White Paper summarising the current status of the theoretical
prediction ahead of the first results from E989.

\section{Lattice formalism and general issues} \label{sec:s2approach}

The hadronic vacuum polarisation contribution can be expressed as a
convolution integral involving the subtracted vacuum polarisation
function, $\hat\Pi(Q^2)\equiv 4\pi^2[\Pi(Q^2)-\Pi(0)]$,
i.e. \cite{Lautrup:1969nc,Blum:2002ii}
\be\label{eq:hvpdef}
  \ahvp=\left(\frac{\alpha}{\pi}\right)^2
  \int_0^{\infty}dQ^2\,f(Q^2)\hat\Pi(Q^2), 
\ee
where the integration is over Euclidean momenta, $\Pi(Q^2)$ is
obtained from the vacuum polarisation tensor
\bea
 \Pi_{\mu\nu}(Q)&=&i\int d^4x\,{\rm e}^{iQ\cdot(x-y)}\,\left\langle
 J_\mu(x) J_\nu(y)\right\rangle \nonumber \\ 
 &\equiv& (Q_\mu
 Q_\nu-\delta_{\mu\nu}Q^2) \Pi(Q^2),
\eea
and $J_\mu=\frac{2}{3}\bar{u}\gamma_\mu u-\frac{1}{3}\bar{d}\gamma_\mu
d- \frac{1}{3}\bar{s}\gamma_\mu s+\ldots$ is the electromagnetic
current. The evaluation of the convolution integral using lattice QCD
data for the correlator is hampered by the fact that the weight
function $f(Q^2)$ is strongly peaked near the muon mass, i.e. for
$|Q|\sim m_\mu$. Since $|Q|$ is inversely proportional to the spatial
lattice extent $L$, the peak region corresponds to lattice volumes
that are much larger than those that can be realised in current
simulations. Thus, there is a lack of lattice data in the region where
the integral receives its dominant contribution.

An alternative formulation in position space, suggested by Bernecker
and Meyer \cite{Bernecker:2011gh}, is known as the time-momentum
representation (TMR)
\be\label{eq:TMRdef}
  \ahvp=\left(\frac{\alpha}{\pi}\right)^2
  \int_0^{\infty}dt\,\tilde{K}(t)G(t),
\ee
\bd
  G(t)\delta_{kl}=-a^3\sum_{\vec{x}}\left\langle J_k(x) J_l(0)
  \right\rangle,
\ed
\bd
  \tilde{K}(t)= 4\pi^2\int_0^{\infty}dQ^2\,f(Q^2)\left[
    t^2-\frac{4}{Q^2}\sin^2\left(\textstyle\frac{1}{2}Qt
    \right)\right],  \nonumber
\ed
where $G(t)$ denotes the spatially summed vector-vector
correlator. Empirically, one finds that the region $t\gtrsim3$\,fm
contributes $\approx3$\% to the integral in \eq{eq:TMRdef}. In order
to determine $\ahvp$ with a precision of 1\% or better, it is clear
that $G(t)$ must be calculated with sufficient accuracy in the
long-distance regime. This, however, is a challenging task, since the
noise-to-signal ratio in $G(t)$ increases exponentially with Euclidean
time $t$.

While both integral representations are equivalent in the sense that
they are formulated in terms of the current-current correlator, the
TMR turns out to be technically simpler: this is due to the fact that
detailed spectral information is available for the vector correlator
in the long-distance regime which is dominated by the isovector
channel. Furthermore, the calculation of quark-disconnected diagrams,
which requires major computational resources (see
Section~\ref{sec:disconn} below) is technically more straightforward.

The main challenges facing lattice QCD regarding a calculation of
$\ahvp$ with a precision that will have a decisive impact on SM tests
are the following: First, the statistical error of the evaluation of
\eq{eq:TMRdef} should lie significantly below 1\%. This implies that
the vector correlator $G(t)$ must be accurately constrained for
$t\gtrsim3$\,fm. The long-distance regime of $G(t)$ receives
significant finite-volume corrections for typical lattice sizes,
i.e. for $m_\pi L\gtrsim4$, which have to be quantified. The isoscalar
part of $G(t)$ requires accurate knowledge of the contributions from
quark-disconnected diagrams which typically exhibit a high level of
statistical noise. Finally, one must include the effects of strong
($m_u\neq m_d$) and electromagnetic isospin breaking corrections. In
the next section we will describe how these issues are addressed in
our recent calculation
\cite{Gerardin:2019rua,Risch:2019xio,Gerardin:2019okl}.

\section{Lattice calculation of $\ahvp$} \label{sec:s3calc}

Our calculations employ the gauge ensembles generated by the CLS
consortium, using $\Nf=2+1$ flavours of O($a$) improved Wilson quarks,
at four different values of the lattice spacing, a range of pion
masses down to the physical one and volumes satisfying $m_\pi L\ge 4$
throughout. For a description of technical details, including the
definition of the vector currents used to compute the correlator
$G(t)$, we refer the reader to
Refs. \cite{Bruno:2014jqa,Gerardin:2019rua,Gerardin:2019okl}.

\subsection{Controlling the infrared regime}

The correlation function $G(t)$ of the electromagnetic current defined
in \eq{eq:TMRdef}, can be split into a sum over the quark-connected
contributions from individual flavours plus the contribution from
quark-disconnected diagrams
\be\label{eq:decompo}
  G(t) = \textstyle\frac{5}{9}G_l(t)+\frac{1}{9}G_s(t)
  +\frac{4}{9}G_c(t)+G_{\rm disc}(t). 
\ee
This expression holds in the isosymmetric case, $m_u=m_d\equiv m_l$,
with the numerical prefactors arising from the electric charges in the
definition of the electromagnetic current. For the following
discussion it is convenient to consider the isospin decomposition of
the vector correlator, i.e.
\be
   G(t)=G^{I=1}(t)+G^{I=0}(t),
\ee
where $G^{I=1}(t)$ and $G^{I=0}(t)$ denote the correlators in the
isovector and isoscalar channels (see Section~2.2.3 in
\cite{Meyer:2018til} for further details). Since the spectral function
in the isoscalar channel vanishes below the 3-pion threshold, one
expects $G(t)$ to be dominated by the lowest-energy state in
$G^{I=1}(t)$ as $t\to\infty$. We have studied three different and
complementary methods to accurately constrain the long-distance regime
of the isovector correlator $G^{I=1}(t)$:

\begin{figure}
\centering
\includegraphics[width=0.45\textwidth,clip]{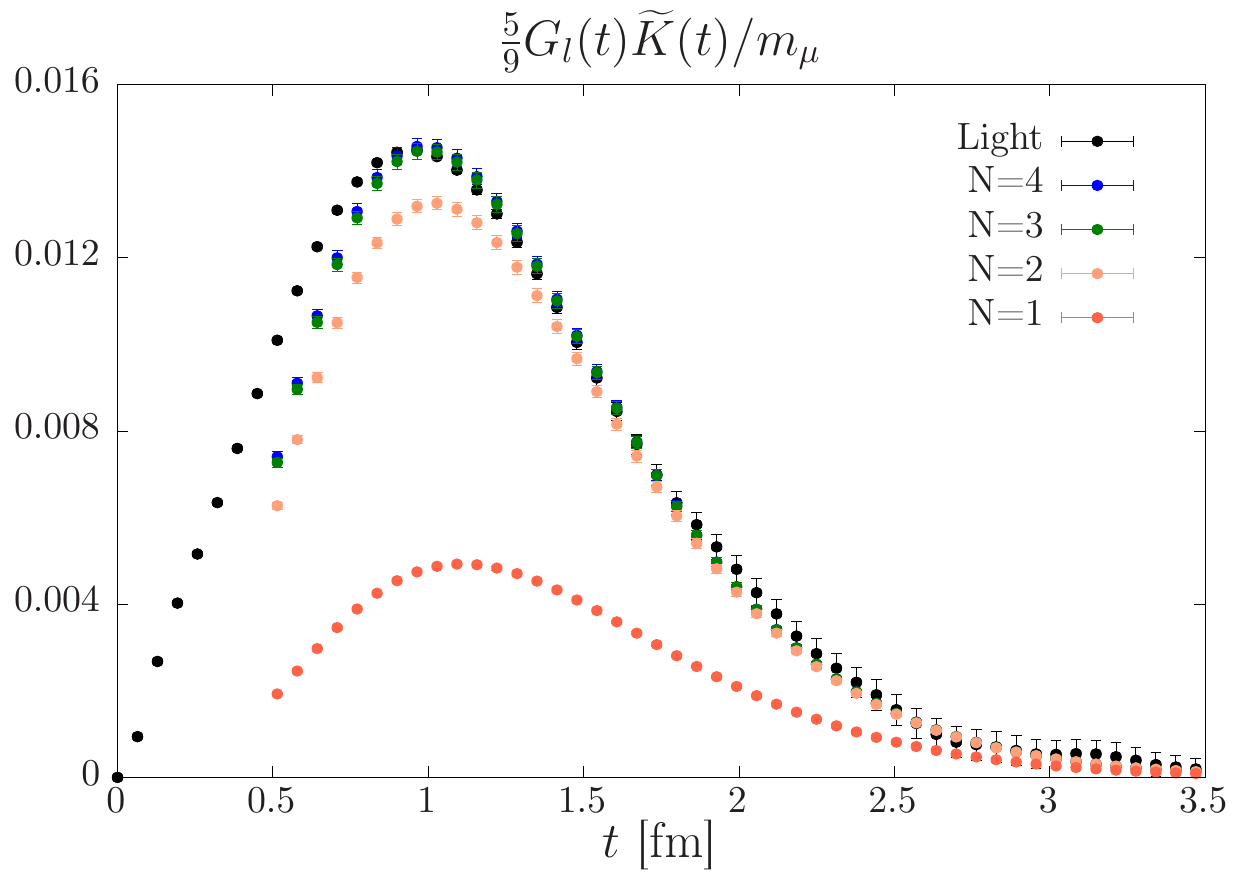}
\caption{The light quark contribution to the integrand of
  \eq{eq:TMRdef} in units of $m_\mu$ (black circles), compared to the
  corresponding reconstruction from the isovector
  contribution. Coloured symbols show the accumulation of up to four
  lowest-lying states in the isovector channel. Data have been
  computed for $m_\pi=200$\,MeV.}
\label{fig:IntegrandD200}
\end{figure}

The first makes use of a dedicated calculation of the energy levels in
the isovector channel, by observing that
\be\label{eq:isovectorasymp}
  G^{I=1}(t)\stackrel{t\to\infty}{=}\sum_n |A_n|^2\,
  {\rm e}^{-\omega_n t},
\ee
where the energies $\omega_n$ are related to the scattering momentum
$k$ and the pion mass via $\omega_n=\sqrt{m_\pi^2+k^2}$
\cite{Luscher:1990ux,Luscher:1991cf}. Figure\,\ref{fig:IntegrandD200}
shows that the 3--4 lowest-lying states saturate the correlator for
distances $t\gtrsim2$\,fm \cite{Gerardin:2019rua}.

\begin{figure*}
\centering
\leavevmode
\includegraphics[width=0.45\textwidth]{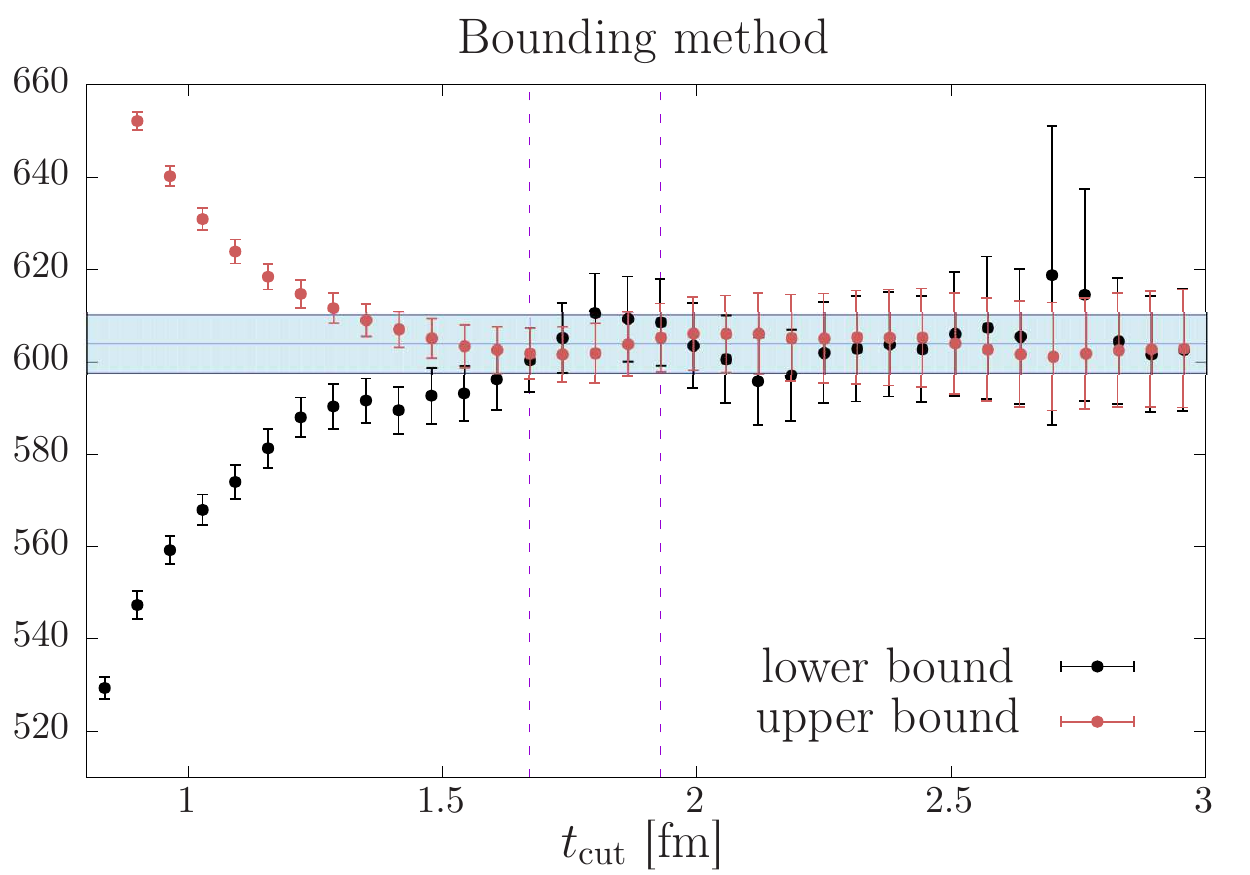}
\hfill
\includegraphics[width=0.45\textwidth]{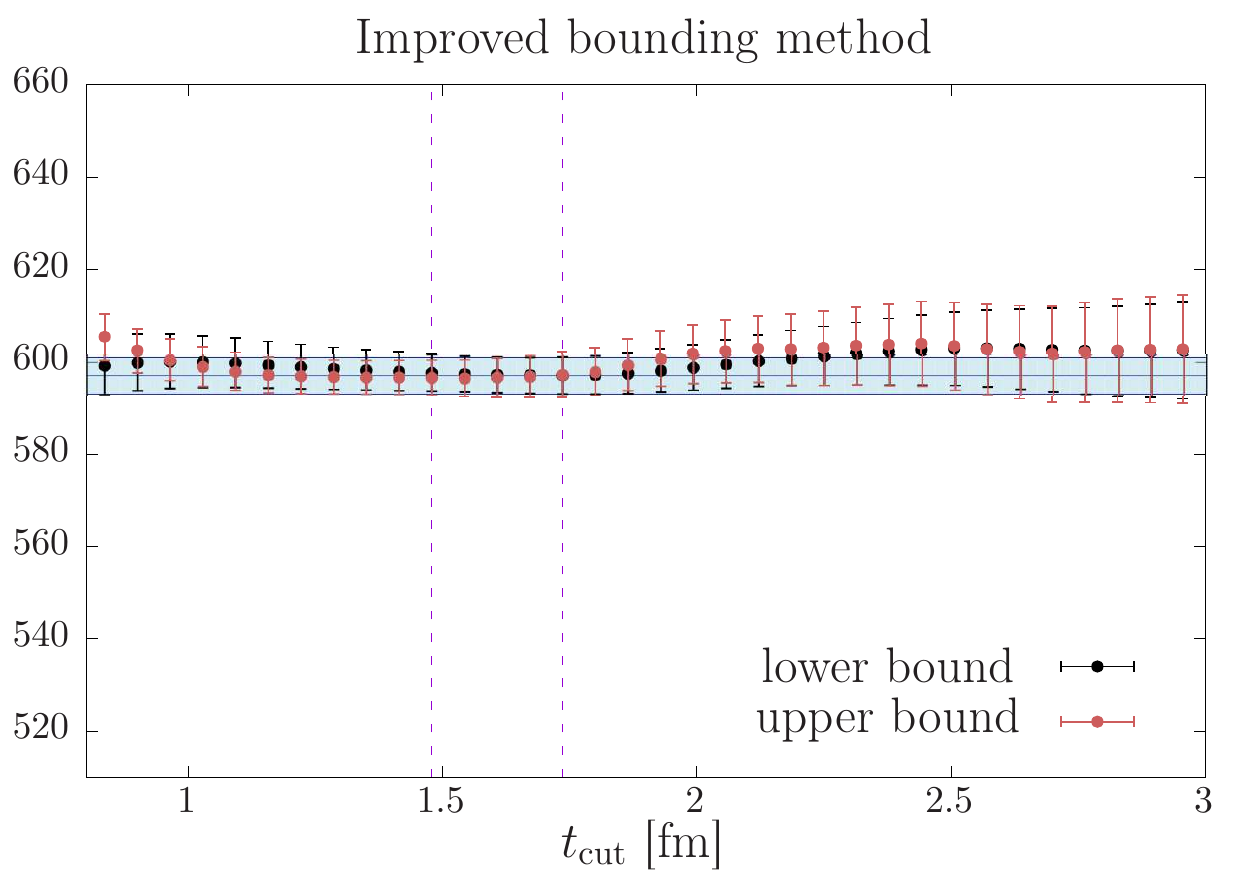}
\caption{Upper and lower bounds for the (connected) light quark
  contribution to $\ahvp$ as a function of $t_{\rm cut}$ computed for
  a pion mass of 200 MeV. The right panel shows the bounds for the
  improved bounding method. Vertical lines indicate the interval over
  which the data are averaged, yielding the horizontal band which is
  identified with the actual estimate.}
\label{fig:bounding}
\end{figure*}

The second procedure, known as the ``bounding method''
\cite{LehnerBounding2016,Borsanyi:2016lpl}, exploits the positivity of
the spectral sum for $G^{I=1}(t)$, which implies that 
\be
   0\leq G(t) \leq G^{I=1}(t_{\rm cut})\,\rme^{-\omega_1(t-t_{\rm
       cut})},
\ee
where $\omega_1$ is the ground state energy, and $t_{\rm cut}$ denotes
a fixed timeslice above which one assumes that the isovector
correlator is well described by the ground state. The bounds on $G(t)$
can be significantly sharpened using the knowledge from the dedicated
calculation of $G^{I=1}(t)$ \cite{Gerardin:2019rua}. By
subtracting the $N-1$ lowest-lying states one defines the
correlator $\tilde{G}(t)$ via
\be
   \tilde{G}(t)\equiv G(t)-\sum_{n=1}^{N-1} |A_n|^2\, 
   \rme^{-\omega_n t}, 
\ee
which satisfies $
%
   0\leq \tilde{G}(t) \leq \tilde{G}(t_{\rm cut})\, 
   \rme^{-\omega_N(t-t_{\rm cut})},
%
$
where $\omega_N$ is the energy of the $N$th state. In
Fig. \ref{fig:bounding} we show a comparison of the ordinary and
improved bounding methods applied to the determination of the
(quark-connected) light quark contribution to $\ahvp$, as a function
of $t_{\rm cut}$. One finds that $\tilde{G}(t)$ saturates the bounds
implied by the improved bounding method already for $t_{\rm
  cut}\gtrsim1.5$\,fm, which, when averaged over a few timeslices,
leads to a stable and statistically precise estimate of $\ahvp$.

The third method to constrain the long-distance regime of the
isovector correlator makes use of the information provided by the
timelike pion form factor $F_\pi(\omega)$ \cite{Meyer:2011um}. It
provides a relation between $F_\pi(\omega)$ and the amplitudes $A_n$
in \eq{eq:isovectorasymp}, which requires knowledge of the
Lellouch-L\"uscher factor \cite{Lellouch:2000pv}, which, in turn, is
determined from the scattering phase shifts
\cite{Luscher:1990ux,Luscher:1991cf} (see \eq{eq:amplitudes} below).
A detailed study of this method in two-flavour QCD can be found in
Ref. \cite{Erben:2019nmx}.

\subsection{Finite-volume effects}

The long-distance regime of the vector correlator is closely related
to the issue of finite-volume corrections. The latter can be
determined by inserting the difference of the vector correlator in
infinite and finite volume, i.e.
\be
 \Delta\ahvp\equiv \left(\frac{\alpha}{\pi}\right)^2
  \int_0^{\infty}dt\,\tilde{K}(t)\left[G(t,\infty)-G(t)\right],
\ee
where $G(t)$ is the correlator in finite volume, and $G(t,\infty)$ is
its infinite-volume counterpart. Realising that the long-distance
regime is most relevant for the estimation of the finite-volume
correction, one can focus on the isovector correlator. In infinite
volume, its spectral representation reads
\be
  G^{I=1}(t,\infty)=\int_0^{\infty}d\omega\,
  \omega^2\rho(\omega^2) \rme^{-\omega|t|},
\ee
where the spectral function is given in terms of the pion form factor
$F_\pi(\omega)$ as
\be
  \rho(\omega^2)=\frac{1}{48\pi^2}\left(1-\frac{4m_\pi^2}{\omega^2}
  \right)^{3/2} |F_\pi(\omega)|^2.
\ee
In finite volume, the expression for the isovector correlator
$G^{I=1}(t)$ has been listed \eq{eq:isovectorasymp}, with amplitudes
given by
\be\label{eq:amplitudes}
  |A_n|^2 = \frac{2k^5}{3\pi\omega^2}
  \frac{|F_\pi(\omega)|^2}{q\phi^\prime(q)+k\delta^\prime(k)}, \quad
  q=kL/2\pi, 
\ee
where $k$ is the scattering momentum. Knowledge of the pion form
factor $F_\pi(\omega)$ allows one to determine the difference
$[G^{I=1}(t,\infty)-G^{I=1}(t)]$ and thus the correction
$\Delta\ahvp$. 

In our evaluation of finite-size effects, we have employed the
Gounaris-Sakurai (GS) representation \cite{Gounaris:1968mw} for
$F_\pi(\omega)$, using the energy levels and amplitudes, $(\omega_n,
A_n)$, as input. Alternatively, we have used $m_\rho$ and the coupling
$g_{\rho\pi\pi}$, taken from a dedicated calculation of
$F_\pi(\omega)$ on a subset of our ensembles \cite{Andersen:2018mau},
to work out $[G^{I=1}(t,\infty)-G^{I=1}(t)]$. We stress that the
somewhat simplistic GS model is used only for the determination of the
finite-size correction but not for the treatment of the long-distance
tail of the correlator itself. In order to test the ability of this
approach to provide an accurate description of finite-size
corrections, we have compared the results for the integrand computed
at a fixed pion mass of 280 MeV for two different volumes,
corresponding to $m_\pi L=3.8$ and 5.8, respectively. Applying the
correction to the data obtained on the smaller volume produces values
that are in excellent agreement with those observed for $m_\pi
L=5.8$. We conclude that finite-size effects are well described by the
GS parameterisation of $F_\pi(\omega)$.

\subsection{Scale setting uncertainty}

At first it may seem surprising that a dimensionless quantity such as
$\ahvp$ should be affected by the lattice scale. However, the kernel
$\tilde{K}(t)$ in \eq{eq:TMRdef} is actually a function of
$(tm_\mu)^2$, and hence it is necessary to express the argument $t/a$
of the vector correlator in units of the muon mass. This requires the
calibration of the lattice spacing $a$. Furthermore, the quark masses
enter the determination of $\ahvp$ indirectly via their feedback onto
the gauge field during the generation of the ensembles. Therefore, the
HVP contribution $\ahvp$ depends on a set of dimensionless ratios
$M_\mu=m_\mu/\Lambda, M_u=m_u/\Lambda,\ldots$ where $\Lambda$ is the
quantity that sets the scale \cite{DellaMorte:2017dyu}:
\be
 \ahvp=\ahvp(M_\mu; M_u, M_d, M_s,\ldots).
\ee
The relative error on the HVP contribution, $\Delta\ahvp/\ahvp$, which
arises from the scale setting uncertainty $\Delta\Lambda/\Lambda$ is
given by
\be
  \frac{\Delta\ahvp}{\ahvp}=\left|
  \frac{M_\mu}{\ahvp}\frac{\partial\ahvp}{\partial M_\mu}+\ldots
  \right| \frac{\Delta\Lambda}{\Lambda},
\ee
where the ellipses represent derivatives with respect to $M_u,
M_d,\ldots$. The logarithmic derivative with respect to $M_\mu$ can
be worked out in terms of an integral representation:
\be
  M_\mu\frac{\partial\ahvp}{\partial M_\mu} =
  -\ahvp+\left(\frac{\alpha}{\pi}\right)^2 \int_0^{\infty}dt\,J(t)G(t),
\ee
where $J(t)=t\tilde{K}^\prime(t)-\tilde{K}(t)$. Using input for the
correlator and the representation of the kernel function described in
Appendix~B of \cite{DellaMorte:2017dyu}, the proportionality factor
between $\Delta\ahvp/\ahvp$ and $\Delta\Lambda/\Lambda$ can be
estimated to be $\approx1.8$. Hence, the scale setting uncertainty
must, at least, be smaller by a factor two compared to the target
precision in the determination of $\ahvp$. The contributions arising
from the quark mass dependence have also been quantified and turn out
to be subdominant \cite{DellaMorte:2017dyu, DellaMorte:2017khn}. In
our calculation, we use the lattice scale determined in
\cite{Bruno:2016plf}, i.e. $\Lambda^{-1}=\sqrt{8t_0}=0.415(4)(2)$\,fm.

\subsection{Isospin breaking}

The effects from strong and electromagnetic isospin breaking must be
taken into account to obtain a lattice QCD estimate for $\ahvp$ whose
precision can compete with that of the data-driven dispersive
approach. One way to include isospin-breaking effects is the so-called
``stochastic method'' which is based on the direct Monte Carlo
evaluation of the path integral in the presence of
electromagnetism. The effects of strong isospin breaking are
incorporated either by choosing different quark masses, $m_u\neq m_d$
or via reweighting techniques. The second method to quantify the
effects of isospin breaking is based on the perturbative expansion of
the path integral in powers of the fine structure constant
$\alpha=e^2/4\pi$ and the mass splitting $m_d-m_u$
\cite{deDivitiis:2011eh,deDivitiis:2013xla}. A detailed comparison of
both methods was performed in \cite{Boyle:2017gzv}.

\begin{figure}
\centering
\includegraphics[width=0.5\textwidth]{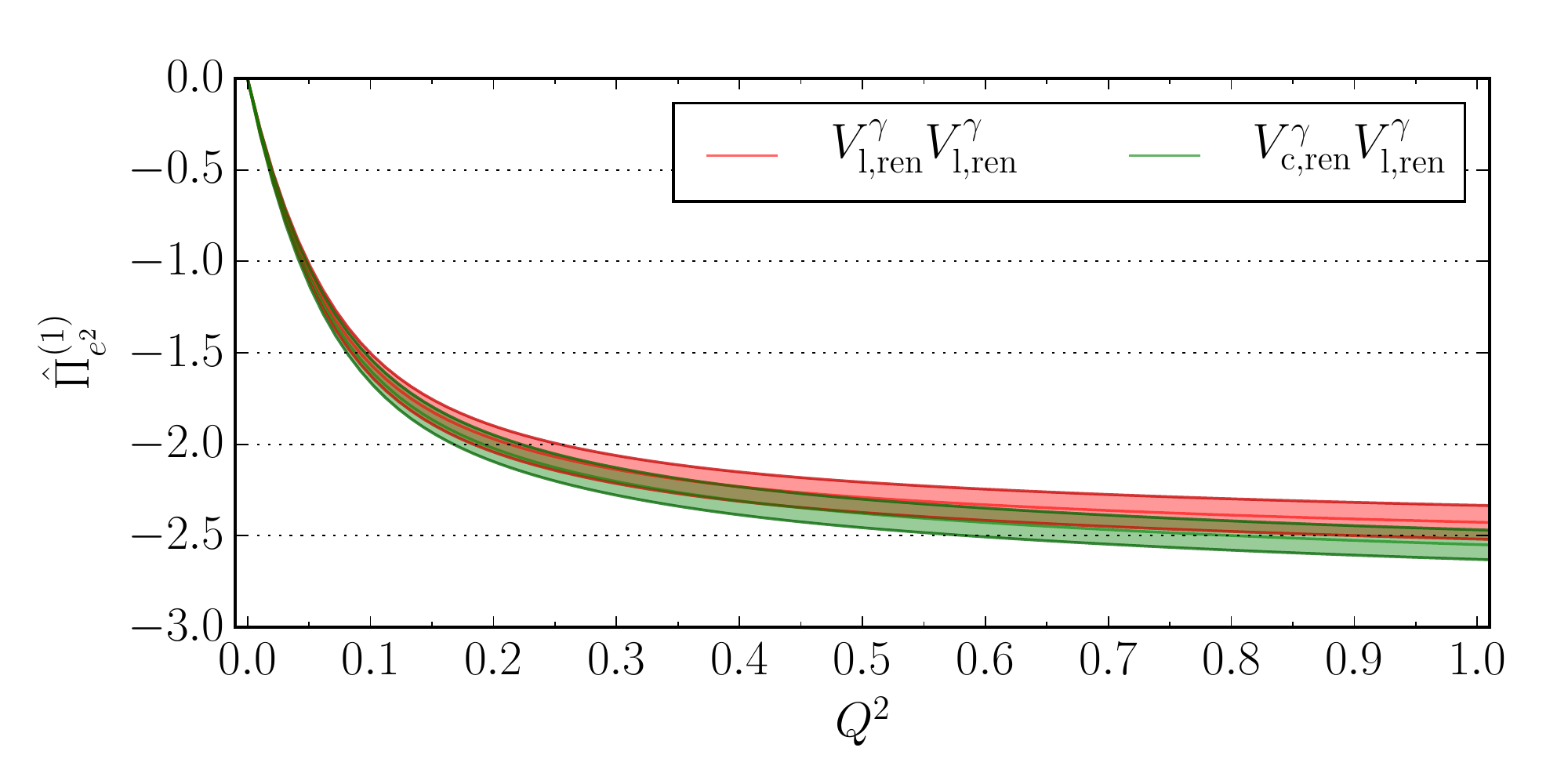}
\vspace{-0.5cm}
\caption{First-order electromagnetic correction to the renormalised
  hadronic vacuum polarisation function, at a pion mass of 350\,MeV
  \cite{Risch:2019xio}. Green and red bands represent different
  discretisations of the vector current.}\label{fig:IB_H102}
\end{figure}

The Mainz group has started to determine isospin-breaking corrections
to $\ahvp$ using the perturbative approach \cite{Risch:2017xxe,
  Risch:2018ozp, Risch:2019xio}. In this way, isospin-breaking effects
can be extracted from additional correlators of the vector current
that involve insertions of the scalar density and explicit photon
propagators (in Coulomb gauge), computed on our existing iso-symmetric
gauge ensembles. Since the photon is a massless unconfined particle,
one expects strong finite-volume effects in the presence of
electromagnetism. Our calculations are based on the $\rm QED_L$
prescription \cite{Hayakawa:2008an} in which the spatial zero modes of
the photon field are set to zero. The necessary adaptation of $\rm
QED_L$ in the presence of open boundary conditions, which are employed
in the generation of most of our gauge ensembles, is discussed in
\cite{Risch:2018ozp}. A list of the additional diagrams for the vector
correlator, as well as first results on the renormalisation factors of
the electromagnetic current and the renormalised hadronic vacuum
polarisation function are shown in \cite{Risch:2019xio} and
Fig.~\ref{fig:IB_H102}.

\begin{figure}
\centering
\includegraphics[width=0.49\textwidth]{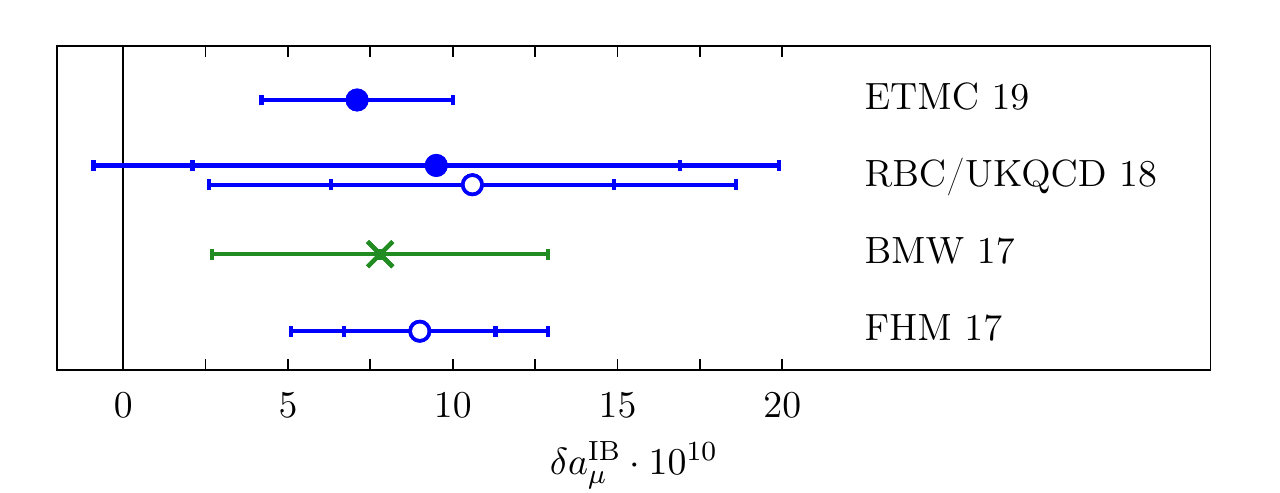}
\vspace{-0.5cm}
\caption{Compilation of recent results for the isospin-breaking
  correction $\delta a_\mu^{\rm IB}$ to the hadronic vacuum
  polarisation, determined by ETMC \cite{Giusti:2019xct},
  RBC/UKQCD\,\cite{Blum:2018mom}, BMW\,\cite{Borsanyi:2017zdw} and
  Fermilab/HPQCD/MILC \cite{Chakraborty:2017tqp}. Solid circles denote
  the full (strong and electromagnetic) isospin-breaking correction,
  while open symbols represent strong isospin breaking only. The
  result marked by the green cross is based on a phenomenological
  estimate.}\label{fig:IB}
\end{figure}

Figure \ref{fig:IB} shows a compilation of recent results for
isospin-breaking corrections obtained by different collaborations. One
finds that the typical size of isospin-breaking effects is
$(5-10)\cdot10^{-10}$, which corresponds to a one-percent correction
to $\ahvp$. While the separation of strong and electromagnetic
isospin-breaking effects is, in principle, scheme-dependent, the
results indicate that the latter are subdominant.

\subsection{Quark-disconnected diagrams \label{sec:disconn}}

While the correlator $G(t)$ contains both quark-connected and
quark-disconnected contributions, the latter are difficult to quantify
with the desired level of precision, owing to the large computational
cost when employing the conventional source method (see, for instance,
the discussion in Section 2.3 of
Ref.\,\cite{Meyer:2018til}). Therefore, stochastic methods are applied
which, despite being numerically quite efficient, introduce stochastic
noise in addition to the statistical fluctuations due to the Monte
Carlo integration over the gauge ensemble. Several methods have been
devised to combat stochastic noise, such as performing a hopping
parameter expansion of the propagator \cite{Bali:2009hu,
  Gulpers:2013uca}, low-mode deflation \cite{Blum:2015you},
hierarchical probing and Hadamard vectors \cite{Stathopoulos:2013aci},
as well as the use of a Lorentz covariant coordinate space formulation
\cite{Meyer:2017hjv} instead of the spatially summed vector
correlator.

An important observation, reported initially in
\cite{Francis:2014hoa}, is the partial cancellation of the stochastic
noise in the difference between the contributions of the light and
strange quarks, $l-s$. This has been exploited in a recently proposed
method \cite{Giusti:2019kff} which uses the so-called one-end-trick
\cite{Foster:1998vw,McNeile:2002fh} in the calculation of the $l-s$
contribution, combined with the hopping parameter expansion applied to
a single heavy quark flavour. Our initial runs show that this method
outperforms hierarchical probing by more than an order of magnitude in
the case of the vector-vector correlator.

\begin{figure}
\centering
\includegraphics[width=0.49\textwidth]{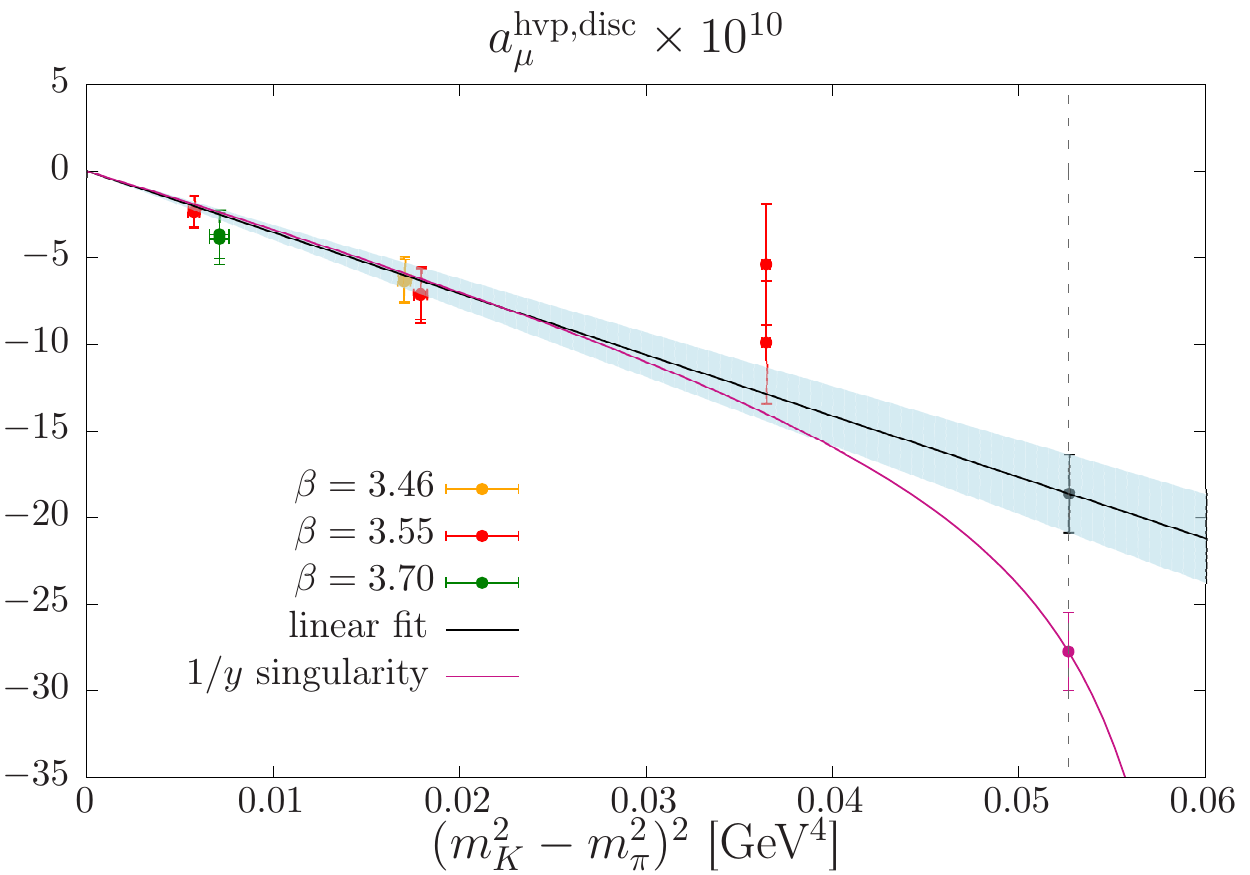}
\vspace{-0.5cm}
\caption{Results for the quark-disconnected contribution $(\ahvp)_{\rm
    disc}$ computed on a subset of our gauge ensembles. Extrapolations
  to the physical mass are performed using two different {\it
    ans\"atze} in the fit.}\label{fig:disconn}
\end{figure}

\begin{figure}
\centering
\includegraphics[width=0.49\textwidth]{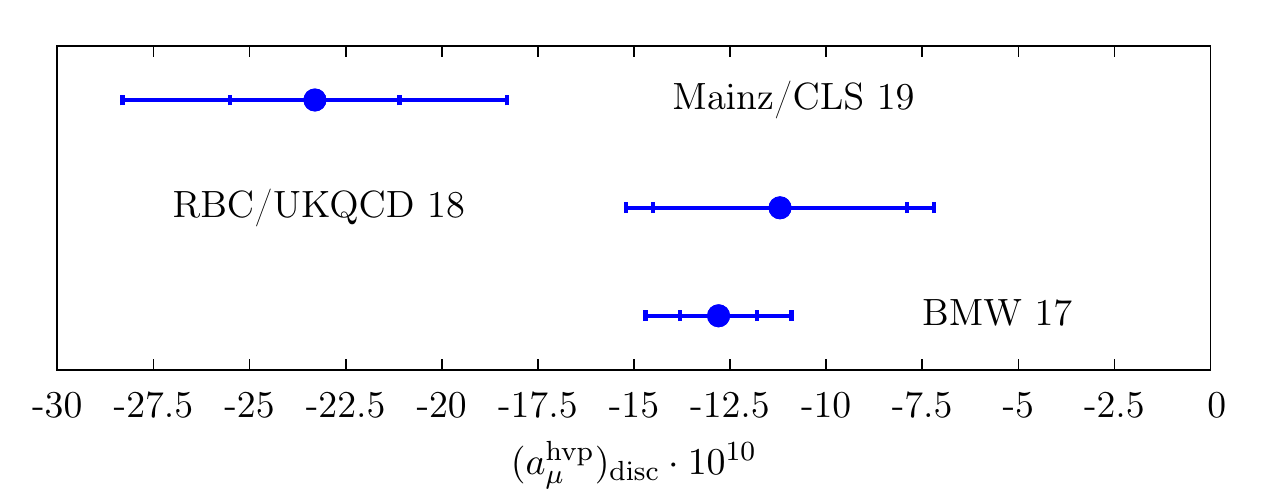}
\vspace{-0.5cm}
\caption{Recent results for the quark-disconnected contribution
  $(\ahvp)_{\rm disc}$ reported by Mainz/CLS \cite{Gerardin:2019rua},
  RBC/UKQCD \cite{Blum:2018mom} and BMW
  \cite{Borsanyi:2017zdw}.}\label{fig:disccomp}
\end{figure}

We have computed the quark-disconnected contribution to $G_{\rm
  disc}(t)$ (see \eq{eq:decompo}) on a subset of our gauge
ensembles. During our calculation we have found that it is difficult
to constrain the long-distance contribution of $G_{\rm disc}(t)$ to
the integrand, since the signal is lost for $t\gtrsim1.5$\,fm. We have
therefore adopted another strategy, which applies the bounding method
to the isoscalar correlator $G^{I=0}(t)$ which contains the
disconnected contribution. Neglecting the contributions from the charm
quark, the positivity of the isoscalar spectral function implies that
the isoscalar correlator satisfies \cite{Gerardin:2019rua}
\be
   0\leq G^{I=0}(t)\leq G^{I=0}(t_{\rm cut})\,
   \rme^{-m_\rho(t-t_{\rm cut})}, 
\ee
where we have assumed $m_\omega\approx m_\rho$, since we do not have a
dedicated calculation of the mass of the $\omega$-meson at our
disposal. The quark-disconnected contribution $(\ahvp)_{\rm disc}$ to
the HVP is then obtained by subtracting the connected contributions of
the light (l) and strange (s) quarks from the isoscalar part, i.e.
\be
   (\ahvp)_{\rm disc} = (\ahvp)^{I=0} 
   -{\textstyle\frac{1}{10}}(\ahvp)^{l}_{\rm con}
   -(\ahvp)^{s}_{\rm con}
\ee
In Fig.\,\ref{fig:disconn} we show the results for $(\ahvp)_{\rm
  disc}$ as a function of $(m_K^2-m_\pi^2)^2$. After extrapolating the
results to the physical mass, we obtain
\be\label{eq:disconn}
   (\ahvp)_{\rm disc} = (-23.2\pm2.2\pm4.5)\cdot 10^{-10},
\ee
where the first error is statistical, and the second is estimated from
the ambiguity in the {\it ansatz} for the chiral extrapolation. A
compilation of recent results for the quark-disconnected contribution
is shown in Fig.\,\ref{fig:disccomp}.

\subsection{Results at the physical point}

\begin{figure*}
\centering
\includegraphics[width=0.7\textwidth]{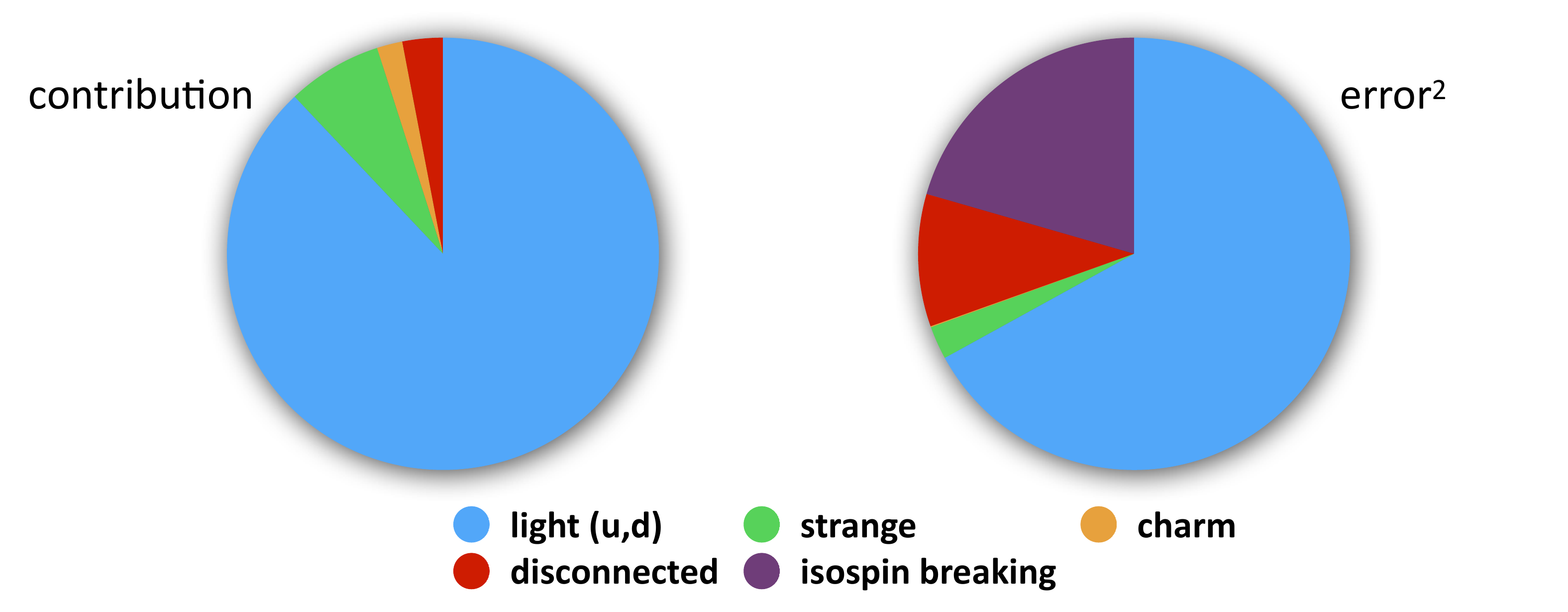}
\caption{Individual contributions to the estimate and variance of our
  result in \eq{eq:final}. While the estimate of $\ahvp$ does not
  contain the correction due to isospin breaking, the size of the
  correction from Ref. \cite{Giusti:2019xct} has been added in
  quadrature to the error.}\label{fig:LatPies}
\end{figure*}

Our results for the hadronic vacuum polarisation contribution have
been obtained on ensembles spanning a range of lattice spacings and
pion masses, including the physical mass. We have subjected our
results for $\ahvp$ for individual flavours to a joint chiral and
continuum extrapolation to the physical point. To this end we have
investigated several {\it ans\"atze}, taking guidance from Chiral
Perturbation Theory \cite{Gerardin:2019rua}. In particular, for the
light quark contribution we have used and compared {\it ans\"atze}
that reproduce the expected singular behaviour as $m_\pi\to0$.

The fact that we have employed two different discretisations for the
vector current \cite{Gerardin:2018kpy} allowed us to extrapolate the
results simultaneously to a common value at the physical point. Only
for the charm quark contribution did we observe large lattice
artefacts for the data extracted from the local-local vector
correlator, so that we decided to extrapolate the results from the
local-conserved correlator, which show much milder discretisation
effects. At the physical point we obtain (in units of $10^{-10}$):
\bea
 & & (\ahvp)^l=674\pm12\pm5, \\
 & & (\ahvp)^s=54.5\pm2.4\pm0.6, \\
 & & (\ahvp)^c=14.66\pm0.45\pm0.06,
\eea
where the first error is statistical and the second gives an estimate
of the systematic uncertainty. The latter is dominated by the form of
the {\it ansatz} for the chiral extrapolation and the scale setting
uncertainty. Our final result is obtained by adding the
quark-disconnected contribution from \eq{eq:disconn} to the sum of the
above estimates. We obtain
\be\label{eq:final}
   \ahvp = (720.0\pm12.4_{\rm stat}\pm9.9_{\rm syst})\cdot10^{-10}.
\ee
Since our calculation of isospin-breaking effects has not been
completed, we have increased the systematic error by adding in
quadrature the result for the full isospin-breaking correction from
Ref.\,\cite{Giusti:2019xct}. The pie charts in Fig.\,\ref{fig:LatPies}
show the individual contributions to the estimate and the variance in
\eq{eq:final}. The total uncertainty of our result is 2.2\%.

%

\section{Comparison, conclusions and outlook}

\begin{figure*}
\centering
\includegraphics[width=0.9\textwidth]{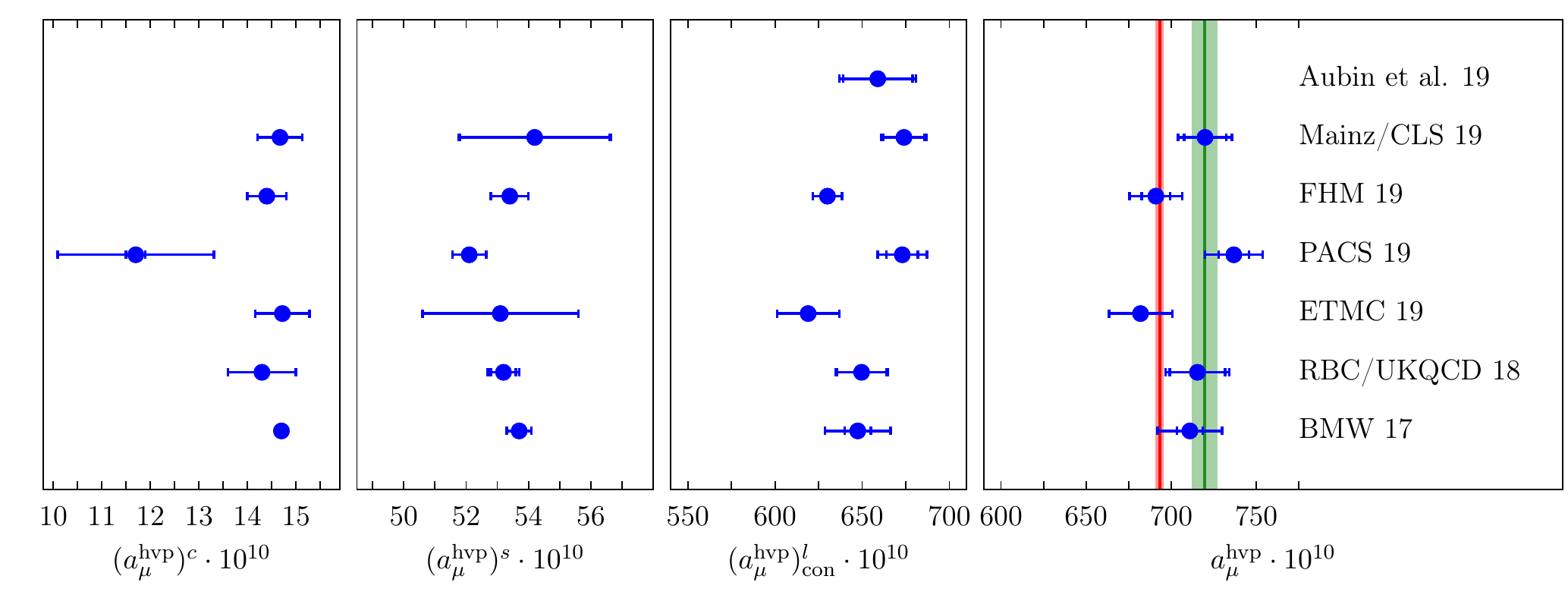}
\vspace{-0.5cm}
\caption{Recent results for the hadronic vacuum polarisation
  contribution from Aubin et al. \cite{Aubin:2019usy}, Mainz/CLS
  \cite{Gerardin:2019rua}, Fermilab/HPQCD/MILC (FHM)
  \cite{Chakraborty:2014mwa, Chakraborty:2016mwy, Davies:2019efs},
  PACS \cite{Shintani:2019wai}, ETMC \cite{Giusti:2017jof,
    Giusti:2018mdh, Giusti:2019xct}, RBC/UKQCD \cite{Blum:2018mom} and
  BMW \cite{Borsanyi:2017zdw}. From left to right the panel represent
  the charm, strange, light (connected) and total contributions to
  $\ahvp$. The vertical red line indicates the result from the
  data-driven analysis of Ref. \cite{Keshavarzi:2018mgv}. The green
  vertical band corresponds to the ``no new physics'' scenario.}
\label{fig:udsc}       
\end{figure*}

Our result \cite{Gerardin:2019rua}, shown in \eq{eq:final}, is
compared to other recent estimates from lattice QCD
\cite{DellaMorte:2017dyu, Borsanyi:2017zdw, Blum:2018mom,
  Giusti:2019xct, Shintani:2019wai, Davies:2019efs, Aubin:2019usy} and
to the data-driven approach in Fig.~\ref{fig:udsc}. 

The two leftmost panels of the figure show that lattice estimates for
the contributions of the strange and charm quarks, $(\ahvp)^s$ and
$(\ahvp)^c$, are in good agreement among different groups, with the
possible exception of the result for $(\ahvp)^c$ by PACS\,19
\cite{Shintani:2019wai}. Given that a range of different
discretisations are employed and that the strange and charm quarks
make only small contributions to the tail of $G(t)$, the observed
consistency can be taken as an indication that lattice artefacts are
under good control.

By contrast, the results for the connected light-quark contribution,
$(\ahvp)^l_{\rm con}$, show a much higher degree of scatter even
though there is a certain degree of correlation among different
estimates: For instance, the calculations by Aubin et
al. \cite{Aubin:2019usy} and Fermilab-HPQCD-MILC (FHM\,19)
\cite{Davies:2019efs} share a subset of the same gauge
ensembles. Furthermore, most groups employ a similar formalism to
quantify finite-volume effects. A weighted average of the results for
$(\ahvp)^l_{\rm con}$ yields $\chi^2/\rm d.o.f.=14.62/6\simeq 2.44$
which illustrates that the scatter among the results is large compared
to the quoted total uncertainty. By contrast, weighted averages of the
estimates for $(\ahvp)^c$ and $(\ahvp)^s$ yield $\chi^2/\rm
d.o.f. \approx 1$ in either case. These observations call for further
scrutiny of the light connected contribution, in particular regarding
the estimation of the tail of $G(t)$ as well as finite-volume effects.

Similar comparisons of isospin-breaking corrections and
quark-disconnected contributions are shown in Figs.\,\ref{fig:IB}
and\,\ref{fig:disccomp}. While the agreement of isospin-breaking
effects among different groups is quite good, the same cannot be said
for the quark-disconnected contribution which, in the case of our own
calculation, turns out to be a factor two larger in magnitude than
what has been reported by other collaborations. Using a new technique
that takes inspiration from Refs.\,\cite{Foster:1998vw,
  McNeile:2002fh, Giusti:2019kff}, we are currently computing the
quark-disconnected contribution with much higher statistical accuracy
and over a wider range of ensembles than reported previously.

The comparison between lattice estimates and the result from
dispersion relations, which is represented by the red vertical
band,\footnote{As a representative example we display the result from
  \cite{Keshavarzi:2018mgv}, noting that a similar result with a
  slightly larger error has been published in \cite{Davier:2019can}.}
shows that the overall precision of lattice calculations is not yet
sufficient to distinguish between the data-driven approach and the
``no new physics'' scenario, which is represented by the green band in
the rightmost panel of Fig.\,\ref{fig:udsc}. The latter corresponds to
the difference between the experimental result for $a_\mu$ and the SM
prediction without the leading-order HVP contribution.

Further efforts are underway to reduce the overall uncertainty of
lattice estimates for both the hadronic vacuum polarisation and
hadronic light-by-light scattering contributions. The status,
prospects and opportunities for lattice calculations applied to the
muon $(g-2)$ will also be discussed in a White Paper which is
currently in preparation.

\medskip
\par\noindent {\bf Acknowledgements.} This work is partly supported by
the Deutsche Forschungsgemeinschaft (DFG, German Research Foundation)
grant HI 2048/1-1 and the DFG-funded Collaborative Research Centre
1044 \emph{The low-energy frontier of the Standard Model}, as well as
by the Cluster of Excellence \emph{Precision Physics, Fundamental
  Interactions, and Structure of Matter} (PRISMA$^+$ EXC 2118/1)
funded by DFG within the German Excellence Strategy (Project ID
39083149). Calculations for this project were performed on HPC
clusters at the Helmholtz-Institut Mainz and at JGU Mainz. Additional
computer time has been allocated through project HMZ21 on the
supercomputer system JUQUEEN at NIC, J\"ulich. The authors gratefully
acknowledge the Gauss Centre for Supercomputing
e.V.\ (www.gauss-centre.eu) for support by providing computing time on
Hazel\,Hen at HLRS Stuttgart under project GCS-HQCD. We are grateful
to our colleagues in the CLS initiative for sharing ensembles.

%
\bibliography{biblio}

\end{document}